\documentclass{llncs}

\usepackage{graphicx}
\usepackage{amsmath,algorithm,algpseudocode,epstopdf,float}
\usepackage[compatibility=false]{caption}
\usepackage{subcaption}

\usepackage{amsfonts,amsmath,amssymb}
\usepackage{algorithmicx}
\usepackage{algpseudocode}
\usepackage{pifont,url}

\algnewcommand{\Inputs}[1]{%
	\State \textbf{Inputs:}
	\Statex \hspace*{\algorithmicindent}\parbox[t]{.8\linewidth}{\raggedright #1}
}
\algnewcommand{\Outputs}[1]{%
	\State \textbf{Outputs:}
	\Statex \hspace*{\algorithmicindent}\parbox[t]{.8\linewidth}{\raggedright #1}
}

\begin{document}

\title{Secure Multi-Party Computation Based Privacy Preserving Extreme Learning Machine Algorithm Over Vertically Distributed Data}

\titlerunning{Secure Multi-Party Computation Based Privacy Preserving Extreme Learning Machine Algorithm Over Vertically Distributed Data}  
%
\author{Ferhat \"{O}zg\"{u}r \c{C}atak}
\authorrunning{Ferhat \"{O}zg\"{u}r \c{C}atak} 
%
\tocauthor{Ferhat Özgür Çatak}
\institute{T\"{U}B\.{I}TAK B\.{I}LGEM, Cyber Security Institute\\
		Kocaeli/Gebze, Turkey,\\
\email{ozgur.catak@tubitak.gov.tr}}

\maketitle 
\begin{abstract}
Especially in the Big Data era, the usage of different classification methods is increasing day by day. The success of these classification methods depends on the effectiveness of learning methods. Extreme learning machine (ELM) classification algorithm is a relatively new learning method built on feed-forward neural-network. ELM classification algorithm is a simple and fast method that can create a model from high-dimensional data sets. Traditional ELM learning algorithm implicitly assumes complete access to whole data set. This is a major privacy concern in most of cases. Sharing of private data (i.e. medical records) is prevented because of security concerns. In this research, we propose an efficient and secure privacy-preserving learning algorithm for ELM classification over data that is vertically partitioned among several parties. The new learning method preserves the privacy on numerical attributes, builds a classification model without sharing private data without disclosing the data of each party to others.

\end{abstract} 
\section{Introduction}
The main purpose of machine learning can be expressed as to find the patterns and summarize the data form of high-dimensional data sets. The classification algorithms \cite{anderson1986machine,ramakrishnan2000database} is one of the most widely used method of machine learning in real-life problems. Data sets used in real-life problems are high dimensional as a result, analysis of them is a complicated process. 

Extreme Learning Machine (ELM) was proposed by \cite{Huang2006489} based on generalized Single-hidden Layer Feed-forward Networks (SLFNs). Main characteristics of ELM are small training time compared to traditional gradient-based learning methods, high generalization property on predicting unseen examples with multi-class labels and parameter free with randomly generated hidden nodes.

Background knowledge attack uses quasi-identifier attributes of a dataset and reduce the possible values of output sensitive information. A well-known example about background knowledge attack is the personal health information about of Massachusetts governor William Weld using a anonymized data set \cite{sweeney2002k}. In order to overcome these types of attacks, various anonymization methods have been developed like \textit{k}-anonymity \cite{sweeney2002k}, \textit{l}-diversity \cite{Machanavajjhala:2007:LDP:1217299.1217302}, \textit{t}-closeness \cite{li2007t}.

Although the anonymization methods are applied to data sets to protect sensitive data, though, the sensitive data is still accessed by an attacker in various ways \cite{ji2014differential}. Also, data anonymization methods are not applicable in some cases. In another scenario, consider the situation when two or more hospitals wants to analyze patient data \cite{lindell2009secure} through collaborative processes that require using each other's databases. In such cases, it is necessary to find a secure training method that can run jointly on private union databases, without revealing or pooling their sensitive data. Privacy-preserving ELM learning systems are one of the methods that the only information learned by the different parties is the output model of learning method.

In this research, we propose a privacy-preserving ELM training model that constructs the global ELM classification model from the distributed data sets in multiple parties. The training data set is vertically partitioned among the parties, and the final distributed model is constructed at an independent party to securely predict the correct label for the new input data. 

The content of this paper is as follows: Related work is reviewed in Section \ref{sec:relworks}. In Section \ref{sec:preliminaries}, ELM, secure multi-party computation and the secure addition are explained. In section \ref{sec:ppelm}, our new  privacy-preserving ELM for vertically partitioned data is proposed. Section \ref{sec:experiments} emprically shows the timing results of our method with different public data sets.
\section{Related Works}\label{sec:relworks}
In this section, we review the existing works that have been developed for different machine learning methods. The major differences between our learning model and existing work are highlighted.

Recently, there has been significant contributions in privacy-preserving machine learning. Secretans et al. \cite{4371189} presents a probabilistic neural network (PNN) model. The PNN is an approximation of the theoretically optimal classifier, known as the Bayesian optimal classifier. There are at least three parties involved in the computation of the secure matrix summation to add the partial class conditional probability vectors together.  Aggarwal et al. \cite{Aggarwal2004}  developed condensation based learning method. They show that an anonymized data closely matches the characteristics of the original data. Samet et al. \cite{Samet:2012:PBE:2350807.2351024}  present new privacy-preserving protocols for both the back-propagation and ELM algorithms among several parties. The protocols are presented for perceptron learning algorithm and applied only single layer models. Oliveria et al. \cite{oliveira2010privacy} proposed methods distort confidential numerical features to protect privacy for clustering analysis. Guang et al. \cite{guang2009privacy} proposed a privacy-preserving back-propagation algorithm for horizontally partitioned databases for multi-party case. They use secure sum in their protocols. Yu et al. \cite{Yu:2006:PSU:1141277.1141415} proposed a privacy-preserving solution for support vector machine classification. Their approach constructs the global SVM classification model from the data distributed at multiple parties, without disclosing the data of each party to others.
\section{Preliminaries}\label{sec:preliminaries}
In this section, we introduce preliminary knowledge of ELM, secure multi-party computation and secure addition briefly. 
\subsection{Extreme learning machine}\label{sec:ELM}
ELM was originally proposed for the single-hidden layer feed-forward neural networks \cite{Huang06extremelearning,1650244,Huang2006489} . Then, ELM was extended to the generalized single-hidden layer feed-forward networks where the hidden layer may not be neuron like \cite{Huang20073056,Huang20083460}. Main advantages of ELM classification algorithm is that ELM can be trained hundred times faster than traditional neural network or support vector machine algorithm since its input weights and hidden node biases are randomly created and output layer weights can be analytically calculated by using a least-squares method \cite{6866146,Huang2008576}. The most noticeable feature of ELM is that its hidden layer parameters are selected randomly.

Given a set of training data $\mathcal{D}=\{(\mathbf{x}_i, y_i)\mid i=1,...,n\},\mathbf{x}_i \in \mathbb{R}^p,\, y_i \in \{1, 2,...,K\}\}$ sampled independently and identically distributed (i.i.d.) from some unknown distribution. The goal of a neural network is to learn a function $f:\mathcal{X} \rightarrow \mathcal{Y}$ where $\mathcal{X}$ is instances and $\mathcal{Y}$ is the set of all possible labels. The output label of an single hidden-layer feed-forward neural networks (SLFNs) with $N$ hidden nodes can be described as
\begin{equation}
\label{eq:slfns}
f_N(\mathbf{x}) = \sum_{i=1}^{N}\beta_iG(\mathbf{a}_i,b_i,\mathbf{x}) , \, \mathbf{x} \in \mathbb{R}^n, \, \mathbf{a}_i \in \mathbb{R}^n
\end{equation}
where $\mathbf{a}_i$ and $b_i$ are the learning parameters of hidden nodes and $\beta_i$ is the weight connecting the $i$th hidden node to the output node. The output function of ELM for generalized SLFNs can be identified by
\begin{equation}
\label{eq:slfnsgen}
f_N(\mathbf{x}) = \sum_{i=1}^{N}\beta_iG(\mathbf{a}_i,b_i,\mathbf{x}) = \mathbf{\beta} \times h(\mathbf{x})
\end{equation}
For the binary classification applications, the decision function of ELM becomes
\begin{equation}
\label{eq:binaryelm}
f_N(\mathbf{x}) = sign\left( \sum_{i=1}^{N}\beta_iG(\mathbf{a}_i,b_i,\mathbf{x}) \right) = sign\left(\mathbf{\beta} \times h(\mathbf{x}) \right)
\end{equation}
Equation \ref{eq:slfnsgen} can be written in another form as 
\begin{equation}
\label{eq:elm}
\mathbf{H}\beta=\mathbf{T}
\end{equation}
$\mathbf{H}$ and $\mathbf{T}$ are respectively hidden layer matrix and output matrix.
\begin{equation}
	\label{eq:elmbeta}
	\beta = \mathbf{H}^\dagger \mathbf{T}
\end{equation}
$\mathbf{H}^\dagger$ is the\textit{ Moore-Penrose generalized inverse of matrix} $\mathbf{H}$. Hidden layer matrix can be described as

\begin{equation}
\label{eq:H}
H(\tilde{a},\tilde{b},\tilde{x})= \begin{bmatrix} G(a_1,b_1,x_1) & \cdots & G(a_L,b_L,x_1) \\ \vdots & \ddots & \vdots \\ G(a_1,b_1,x_N) & \cdots & G(a_L,b_L,x_N) \end{bmatrix}_{N \times L}
\end{equation}
where $\tilde{a}=a_1,...,a_L$, $\tilde{b}=b_1,...,b_L$, $\tilde{x}=x_1,...,x_N$. Output matrix can be described as
\begin{equation}
\label{eq:elmoutput}
T= \begin{bmatrix} t_1  \hdots  t_N \end{bmatrix}^T
\end{equation}
The hidden nodes of SLFNs can be randomly generated. They can be independent of the training data.
\subsection{Secure Multi Party Computation}\label{sec:SMPC}
\begin{figure}
	\centering
	\includegraphics[width=0.6\linewidth]{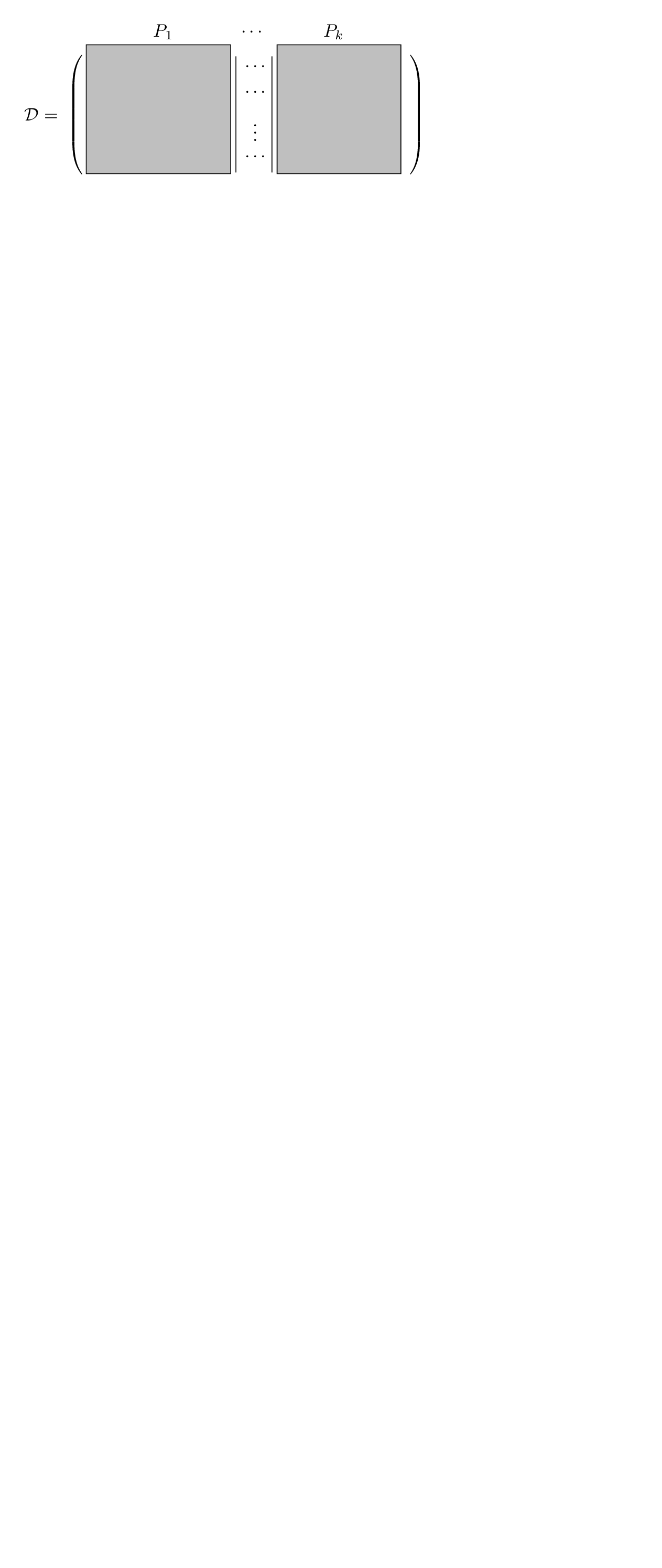}
	\caption{Vertically partitioned data set $\mathcal{D}$. }
	\label{fig:partititon}
\end{figure}

In vertically partitioned data, each party holds different attributes of same data set. Let's have $n$ input instances, $\mathcal{D} = \left\{ (\mathbf{x}_i, y_i)\mid\mathbf{x}_i \in \mathbb{R}^p,\, y_i \in \mathbb{R}\right\}_{i=1}^n$. The partition strategy is shown in Figure \ref{fig:partititon}. 
\subsubsection{Secure Multi-Party Addition}
In secure multi-party addition (SMA), each party, $P_i$, has a private local value, $x_i$. At the end of the computation, we obtain the sum, $x=\sum_{i=0}^{k-1}$. For this works, we applied the Yu et al. \cite{yu2006privacy} secure addition procedure. Their approach is a generalization of the existing works \cite{sweeney2004multiparty} that uses secure communication and trusted party. Canonical order based , $P_0, \cdots, P_{k-1}$, protocol is applied. The SMA method is show in Algorithm \ref{alg:sma}. This protocol calculates the required sum in secure manner.
\begin{algorithm}[h]
	\caption{Secure multi-party addition}
	\label{alg:sma}
	\begin{algorithmic}[1]
		\Procedure{SMA}{$\mathbf{P}$}
			\State $P_0: R \gets rand(\mathcal{F})$ \Comment{$P_0$ randomly chooses a number $R$}
			\State $V \gets R + x_0\mod{\mathcal{F}}$
			\State $P_0$ sends $V$ to node $P_1$
 			\For{$i=1,\cdots,k-1$ }
 				\State $P_i$ receives $V=R+\sum_{j=0}^{i-1}{x_j\mod{\mathcal{F}}}$
 				\State $P_i$ computes $V=\left( R+\sum_{j=1}^{i}{x_j\mod{\mathcal{F}}} \right) = \left( (x_i + V)\mod{\mathcal{F}} \right) $
 				\State $P_i$ sends $V$ to node $P_{i+1}$
			\EndFor
			\State $P_0: V \gets \left( V - R \right) = \left( V - R\mod{\mathcal{F}} \right)$ \Comment{Actual addition result}
		\EndProcedure
\end{algorithmic}
\end{algorithm}
\section{Privacy-preserving ELM over vertically partitioned data}\label{sec:ppelm}
The data set that one wants to find a classifier for consists of $m$ instances in $n$-dimensional space is shown with $\mathcal{D} \in \mathbb{R}^{m \times n}$. Each instances of the data set has values for $n$ features. Matrix $\mathcal{D}$ is vertically partitioned into $i$ parties of $P_0, P_1, \cdots, P_i$  and each features of instances owned by a party that is private shown as Figure \ref{fig:partititon} illustrates. As shown in Equation \ref{eq:H}, at second stage of ELM learning, hidden layer output matrix, $\mathbf{H}$ is calculated using randomly assigned hidden node parameters $\mathbf{w}$, $\mathbf{b}$ and $\beta$. ELM calculates the matrix $\mathbf{H}$ and output weight vector, $\beta$, is obtained by multiplying $\mathbf{H}$ and $\mathbf{T}$. Each member of $\mathbf{H}$ is computed with an activation function $g$ such that $G(\mathbf{w}_i, \mathbf{x}_i, b_i) = g(\mathbf{x}_i \cdot \mathbf{w}_i + b_i)$ for sigmoid or $G(\mathbf{w}_i, \mathbf{x}_i, b_i) = g(b_i - || \mathbf{x}_i - \mathbf{w}_i ||)$ for radial based functions. An $(i,j)^{th}$ element of $\mathbf{H}$ is 
\begin{equation}
	G(\mathbf{w}_i, \mathbf{x}_i, b_i) = sign(\mathbf{x}_i \cdot \mathbf{w}_i +b_i )
\end{equation}
where $\mathbf{x}_i$ is the $i$th instances of data set and $\mathbf{w}_i$ is hidden node input weight of $i$th instance and $\mathbf{x}_i, \mathbf{w}_i \in \mathbb{R}^{n} $.

Let $\mathbf{x}_i^1, \cdots \mathbf{x}_i^k$ be vertically partitioned vectors of input instance $\mathbf{x}_i$ and $\mathbf{w}_j^1, \cdots \mathbf{w}_j^k$ be vertically partitioned vectors of $j$th hidden node input weight $\mathbf{w}_j$, $b_j^0, \cdots , b_j^k$ be $j$th node input bias over $k$ different parties. Then the output of $j$th input node with $i$th instance of input data set using $k$ sites is
\begin{equation}
	\label{eq:secH}
	sign(\mathbf{x}_i \cdot \mathbf{w}_j +b_j ) = sign\left( \left(\mathbf{x}_i^0 \cdot \mathbf{w}_j^0 + b_j^0) + \cdots  + (\mathbf{x}_j^k \cdot \mathbf{w}_j^k + b_j^k \right) \right)
\end{equation}
From Equation \ref{eq:secH}, calculation of hidden layer output matrix, $\mathbf{H}$ can be decomposed into $k$ different parties using secure sum of matrices, such that
\begin{equation}
	\mathbf{H} = sign \left( \mathbf{T}_1 + \cdots + \mathbf{T}_k \right)
\end{equation}
where
\begin{equation}
	\mathbf{T}_i = \begin{bmatrix} \left(\mathbf{x}_1^i \cdot \mathbf{w}_1^i + b_1^i\right)  & \cdots & \left(\mathbf{x}_1^i \cdot \mathbf{w}_L^i + b_L^i\right)  \\ \vdots & \ddots & \vdots \\ \left(\mathbf{x}_N^i \cdot \mathbf{w}_1^i + b_1^i\right) & \cdots & \left(\mathbf{x}_L^i \cdot \mathbf{w}_N^i + b_N^i\right) \end{bmatrix}_{N \times L}
\end{equation}
\subsubsection{Privacy Preserving ELM Algorithm}
Let $\mathcal{D} \in \mathbb{R}^{M \times N}$, and the number of input layer size be $L$, the number of parties be $k$, then the our training model becomes:
\begin{enumerate}
	\item Master party creates weight matrix, $\mathbf{W} \in \mathbb{R}^{L \times N}$
	\item Master party distributes partition $\mathbf{W}$ with same feature size for each parties.
	\item Party $P_0$ creates a random matrix, 
	$ \mathbf{R} =\begin{bmatrix} rand_{1,1}(\mathcal{F}) & \cdots  & rand_{1,L}(\mathcal{F}) \\ \vdots & \ddots & \vdots \\ rand_{N,1}(\mathcal{F}) & \cdots  & rand_{N,L}(\mathcal{F})  \end{bmatrix}_{N \times L}$
	\item Party $P_0$ creates perturbated output,
	$\mathbf{V} = \mathbf{R} + \begin{bmatrix} \left(\mathbf{x}_1^0 \cdot \mathbf{w}_1^0 + b_1^0\right)  & \cdots & \left(\mathbf{x}_1^0 \cdot \mathbf{w}_L^0 + b_L^0\right)  \\ \vdots & \ddots & \vdots \\ \left(\mathbf{x}_N^0 \cdot \mathbf{w}_1^0 + b_1^0\right) & \cdots & \left(\mathbf{x}_N^0 \cdot \mathbf{w}_L^0 + b_L^0\right) \end{bmatrix}$
	\item for $i=1,\cdots,k-1$ 
	\begin{itemize}
		\item $P_i$ computes 
		$\mathbf{V} = \mathbf{V} + \begin{bmatrix} \left(\mathbf{x}_1^i \cdot \mathbf{w}_1^i + b_1^i\right)  & \cdots & \left(\mathbf{x}_1^i \cdot \mathbf{w}_L^i + b_L^i\right)  \\ \vdots & \ddots & \vdots \\ \left(\mathbf{x}_N^i \cdot \mathbf{w}_1^i + b_1^i\right) & \cdots & \left(\mathbf{x}_L^i \cdot \mathbf{w}_N^i + b_N^i\right) \end{bmatrix}$
		\item $P_i$ sends $\mathbf{V}$ to $P_{i+1}$
	\end{itemize}
	\item $P_0$ subtracts random matrix, $\mathbf{R}$, from the received matrix $\mathbf{V}$.
	$\mathbf{H} = \left( \mathbf{V} - \mathbf{R} \right) \mod{\mathcal{F}} $
	\item Hidden layer node weight vector, $\beta$, is calculated. 
	$\beta = \mathbf{H}^\dagger \cdot \mathbf{T}$
\end{enumerate}
\section{Experiments}\label{sec:experiments}
In this section, we perform experiments on real-world data sets from the public available data set repositories. Public data sets are used to evaluate the proposed learning method. Classification models of each data set are compared for accuracy results without using secure multi-party computation. 

\textit{\textbf{Experimental setup}}: In this section, our approach is applied to six different data sets to verify model affectivity  and efficiency. The data sets are summarized in Table \ref{tbl:dslist}, including australian, colon-cancer, diabetes, duke, heart, ionosphere. 
\begin{table}[h!]
	\caption{Description of the testing data sets used in the experiments.}
	\label{tbl:dslist}
	\begin{center}
		\begin{tabular}{|c||r|r|c|c|}
			\hline Data set & \#Train & \#Classes & \#Attributes \\ 
			\hline \hline australian \cite{quinlan1987simplifying} & 690 & 2 & 14 \\ 
			\hline colon-cancer \cite{alon1999broad} & 62 & 2 & 2,000 \\ 
			\hline diabetes \cite{smith1988using} & 768 & 2  & 8 \\ 
			\hline duke breast cancer \cite{west2001predicting} & 44 & 2 & 7,129 \\
			\hline heart \cite{statlogheart} & 270 & 2 & 13 \\
			\hline ionosphere \cite{Sigillito89} & 351 & 2 & 34 \\
			\hline 
		\end{tabular} 
	\end{center}
\end{table}
For each data set in Table \ref{tbl:dslist}, we vary number of party size, $k$ from $2$ to number of feature, $n$, of the data set. For instance, when our party size is three, $k=3$, and attribute size fourteen, $n=14$, then the first two party have 5 attributes, and last party has 4 attributes.

\textit{\textbf{Simulation Results}}: The accuracy of secure multi-party computation based ELM is exactly same for the traditional ELM training algorithm.
\begin{figure}[h!]
	\centering
	\includegraphics[width=0.7\linewidth]{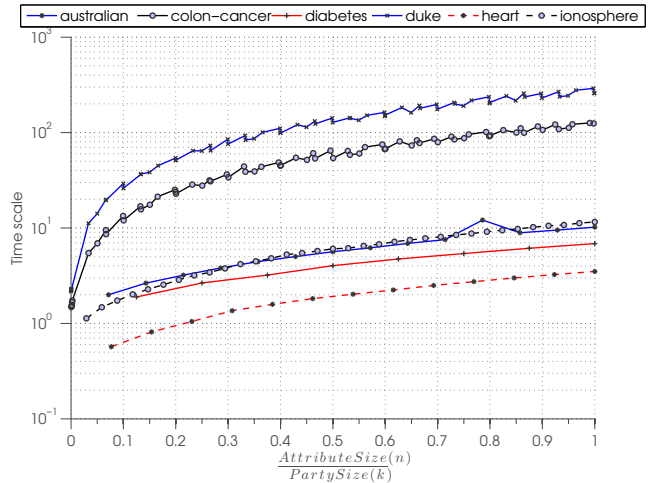}
	\caption{Vertically partitioned data set $\mathcal{D}$. }
	\label{fig:results}
\end{figure}
Figure \ref{fig:results} shows results of our simulations. As shown in figure, time scale becomes its steady state position when number of parties, $k$, moves closer to number of attributes, $k$. 
\section{Conclusion and Future Works}
ELM learning algorithm, a new method compared to other classification algorithms. ELM outperforms traditional Single Layer Feed-forward Neural-networks and Support Vector Machines for big data \cite{6733226}. The ELM is applied in many fields. Almost, in all fields that ELM is applied (i.e. medical records, business, government), privacy is a major concern. 

A new privacy-preserving learning model is proposed for ELM in vertically partitioned data in multi-party partitioning without sharing the data of each site to the others. In order to save the privacy of input data set, master party divides weight vector, and each party calculates the activation function result with its data and weight vector. Extending the privacy-preserving ELM to horizontally distributed data set is a future work for this approach. 
\bibliographystyle{splncs}
\bibliography{references}

\begin{thebibliography}{10}

\bibitem{anderson1986machine}
Anderson, J.R., Michalski, R.S., Carbonell, J.G., Mitchell, T.M.:
\newblock Machine learning: An artificial intelligence approach. Volume~2.
\newblock Morgan Kaufmann (1986)

\bibitem{ramakrishnan2000database}
Ramakrishnan, R., Gehrke, J.:
\newblock Database management systems.
\newblock Osborne/McGraw-Hill (2000)

\bibitem{Huang2006489}
Huang, G.B., Zhu, Q.Y., Siew, C.K.:
\newblock Extreme learning machine: Theory and applications.
\newblock Neurocomputing \textbf{70}(1–3) (2006)  489 -- 501

\bibitem{sweeney2002k}
Sweeney, L.:
\newblock k-anonymity: A model for protecting privacy.
\newblock International Journal of Uncertainty, Fuzziness and Knowledge-Based
  Systems \textbf{10}(05) (2002)  557--570

\bibitem{Machanavajjhala:2007:LDP:1217299.1217302}
Machanavajjhala, A., Kifer, D., Gehrke, J., Venkitasubramaniam, M.:
\newblock L-diversity: Privacy beyond k-anonymity.
\newblock ACM Trans. Knowl. Discov. Data \textbf{1}(1) (March 2007)

\bibitem{li2007t}
Li, N., Li, T., Venkatasubramanian, S.:
\newblock t-closeness: Privacy beyond k-anonymity and l-diversity.
\newblock In: Data Engineering, 2007. ICDE 2007. IEEE 23rd International
  Conference on, IEEE (2007)  106--115

\bibitem{ji2014differential}
Ji, Z., Lipton, Z.C., Elkan, C.:
\newblock Differential privacy and machine learning: a survey and review.
\newblock arXiv preprint arXiv:1412.7584 (2014)

\bibitem{lindell2009secure}
Lindell, Y., Pinkas, B.:
\newblock Secure multiparty computation for privacy-preserving data mining.
\newblock Journal of Privacy and Confidentiality \textbf{1}(1) (2009) ~5

\bibitem{4371189}
Secretan, J., Georgiopoulos, M., Castro, J.:
\newblock A privacy preserving probabilistic neural network for horizontally
  partitioned databases.
\newblock In: Neural Networks, 2007. IJCNN 2007. (Aug 2007)  1554--1559

\bibitem{Aggarwal2004}
Aggarwal, C., Yu, P.:
\newblock A condensation approach to privacy preserving data mining.
\newblock In Bertino, E., Christodoulakis, S., Plexousakis, D., Christophides,
  V., Koubarakis, M., Böhm, K., Ferrari, E., eds.: Advances in Database
  Technology - EDBT 2004. Volume 2992 of Lecture Notes in Computer Science.
\newblock Springer Berlin Heidelberg (2004)  183--199

\bibitem{Samet:2012:PBE:2350807.2351024}
Samet, S., Miri, A.:
\newblock Privacy-preserving back-propagation and extreme learning machine
  algorithms.
\newblock Data Knowl. Eng. \textbf{79-80} (September 2012)  40--61

\bibitem{oliveira2010privacy}
Oliveira, S.R., Zaiane, O.R.:
\newblock Privacy preserving clustering by data transformation.
\newblock Journal of Information and Data Management \textbf{1}(1) (2010) ~37

\bibitem{guang2009privacy}
Guang, L., Ya-Dong, W., Xiao-Hong, S.:
\newblock A privacy preserving neural network learning algorithm for
  horizontally partitioned databases.
\newblock Inform. Technol. J \textbf{9} (2009)  1--10

\bibitem{Yu:2006:PSU:1141277.1141415}
Yu, H., Jiang, X., Vaidya, J.:
\newblock Privacy-preserving svm using nonlinear kernels on horizontally
  partitioned data.
\newblock In: Proceedings of the 2006 ACM Symposium on Applied Computing. SAC
  '06, New York, NY, USA, ACM (2006)  603--610

\bibitem{Huang06extremelearning}
bin Huang, G., yu~Zhu, Q., kheong Siew, C.:
\newblock Extreme learning machine: A new learning scheme of feedforward neural
  networks.
\newblock In: IN PROC. INT. JOINT CONF. NEURAL NETW. (2006)  985--990

\bibitem{1650244}
Huang, G.B., Chen, L., Siew, C.K.:
\newblock Universal approximation using incremental constructive feedforward
  networks with random hidden nodes.
\newblock Neural Networks, IEEE Transactions on \textbf{17}(4) (July 2006)
  879--892

\bibitem{Huang20073056}
Huang, G.B., Chen, L.:
\newblock Convex incremental extreme learning machine.
\newblock Neurocomputing \textbf{70}(16–18) (2007)  3056 -- 3062

\bibitem{Huang20083460}
Huang, G.B., Chen, L.:
\newblock Enhanced random search based incremental extreme learning machine.
\newblock Neurocomputing \textbf{71}(16–18) (2008)  3460 -- 3468

\bibitem{6866146}
Tang, J., Deng, C., Huang, G.B., Zhao, B.:
\newblock Compressed-domain ship detection on spaceborne optical image using
  deep neural network and extreme learning machine.
\newblock Geoscience and Remote Sensing, IEEE Transactions on \textbf{53}(3)
  (March 2015)  1174--1185

\bibitem{Huang2008576}
Huang, G.B., Li, M.B., Chen, L., Siew, C.K.:
\newblock Incremental extreme learning machine with fully complex hidden nodes.
\newblock Neurocomputing \textbf{71}(4–6) (2008)  576 -- 583

\bibitem{yu2006privacy}
Yu, H., Vaidya, J., Jiang, X.:
\newblock Privacy-preserving svm classification on vertically partitioned data.
\newblock In: Advances in Knowledge Discovery and Data Mining.
\newblock Springer (2006)  647--656

\bibitem{sweeney2004multiparty}
Sweeney, L., Shamos, M.:
\newblock Multiparty computation for randomly ordering players and making
  random selections.
\newblock Technical report, Carnegie Mellon University (2004)

\bibitem{quinlan1987simplifying}
Quinlan, J.R.:
\newblock Simplifying decision trees.
\newblock International journal of man-machine studies \textbf{27}(3) (1987)
  221--234

\bibitem{alon1999broad}
Alon, U., Barkai, N., Notterman, D.A., Gish, K., Ybarra, S., Mack, D., Levine,
  A.J.:
\newblock Broad patterns of gene expression revealed by clustering analysis of
  tumor and normal colon tissues probed by oligonucleotide arrays.
\newblock Proceedings of the National Academy of Sciences \textbf{96}(12)
  (1999)  6745--6750

\bibitem{smith1988using}
Smith, J.W., Everhart, J., Dickson, W., Knowler, W., Johannes, R.:
\newblock Using the adap learning algorithm to forecast the onset of diabetes
  mellitus.
\newblock In: Proceedings of the Annual Symposium on Computer Application in
  Medical Care, American Medical Informatics Association (1988)  261

\bibitem{west2001predicting}
West, M., Blanchette, C., Dressman, H., Huang, E., Ishida, S., Spang, R.,
  Zuzan, H., Olson, J.A., Marks, J.R., Nevins, J.R.:
\newblock Predicting the clinical status of human breast cancer by using gene
  expression profiles.
\newblock Proceedings of the National Academy of Sciences \textbf{98}(20)
  (2001)  11462--11467

\bibitem{statlogheart}
UCI:
\newblock Statlog (heart) data set.
\newblock \url{https://archive.ics.uci.edu/ml/datasets/Statlog+(Heart)} (2015)

\bibitem{Sigillito89}
Sigillito, V.G., Wing, S.P., Hutton, L.V., Baker, K.B.:
\newblock {Classification of radar returns from the ionosphere using neural
  networks}.
\newblock Johns Hopkins APL Technical Digest \textbf{10} (1989)  262--266

\bibitem{6733226}
Cambria, E., Huang, G.B.e.a.:
\newblock Extreme learning machines [trends controversies].
\newblock Intelligent Systems, IEEE \textbf{28}(6) (Nov 2013)  30--59

\end{thebibliography}
\end{document}